**Comment on "Above, Below, and In-Between the Two Glass Transitions of Ultrathin Free-Standing Polystyrene Films: Thermal Expansion Coefficient and Physical Aging" by Justin E. Pye and Connie B. Roth, *J. Polym. Sci., Part B Polym.Phys.* 2015, *53*, 64-75.**

K.L. Ngai[a], S. Capaccioli[b] and D. Prevosto[b]

[a]*CNR-IPCF, Largo Bruno Pontecorvo 3 ,I-56127, Pisa, Italy*

[b]*Dipartimento di Fisica, Università di Pisa, Largo Bruno Pontecorvo 3 ,I-56127, Pisa, Italy*

Pye and Roth (PR)[1] used ellipsometry to measure the thermal expansion of freestanding ultrathin polystyrene films over an extended temperature range. They found the first experimental evidence of the presence of two separate mechanisms acting simultaneously on high molecular weight (MW) PS freestanding thin films, giving rise to two distinctly different reduced glass transition temperatures. The upper transition at $T_{gu}(h)$ with lesser reduction from the bulk $T_g$ has no MW-dependence. The lower transition at $T_{gl}(h)$ is strongly MW-dependent, and $T_{gl}(h)$ shows the linear decrease on decreasing $h$ seen first by Dalnoki-Veress et al.[2], and accepted so far by the community[3] as the sole glass transition temperature in high MW freestanding films by ellipsometry[2], Brillouin light scattering[4], fluorescence measurements[5], photon correlation spectroscopy[6], and dielectric relaxation[7,8].

For all the thin films, PR observe a much stronger and broader upper $T_{gu}(h)$. Based on the change in slope, i.e., thermal expansion coefficient of the films, they determines that upon cooling from the melt to the glass, the majority of the film (~90%) solidifies at the upper transition temperature $T_{gu}(h)$, while only a small fraction (~10%) remains mobile to much lower temperatures, solidifying at the lower transition temperature $T_{gl}(h)$. At the time of publication of the paper[1], PR could not find any theory capable to capture the two different $T_{gu}(h)$ and $T_{gl}$ and $h$-dependences. The only conclusion they can made is that their results indicate two separate mechanisms propagate enhanced mobility from the free surface into the film simultaneously, but cannot provide information as to the location of the faster and minor population within the film. They believed[1] the upper and the lower transitions are genuine glass transitions both manifested by the segmental α-relaxation of PS in different regions of the thin film. This belief was reaffirmed in a recent publication[9]. However, we had pointed out[10] that this scenario contradicts the experimental fact[1] of rapid increase of the difference, $\Delta T_{gul} \equiv T_{gu}(h) - T_{gl}$, with decrease $h$. On thinning the film, the two regions where the upper and lower transitions both caused by the segmental α-relaxation tend to merge. Hence the distinction between the two mechanisms of the upper and lower transitions acting on the segmental α-relaxation at different regions of the film has to gradually disappear, and $\Delta T_{gul}$ must decrease instead the observed rapid increase [1]. This experimental fact poses a problem for the assertion of PR that both the upper and lower glass transitions originates from the segmental α-relaxation. In this Comment, we present a simpler and transparent argument to show that the interpretation of PR contradicts previous experimental findings and hence is invalid.

Besides ellipsometry, glass transition in freestanding high molecular weight polystyrene thin films of thickness $h$ have been studied by dielectric relaxation [7,8], Brillouin light scattering [4], fluorescence measurements [5], photon correlation spectroscopy (PCS) [6]. These techniques are sensitive probes of the segmental α-relaxation, and would not miss detecting the much stronger upper transition over ~90% of the film together with the weaker lower transition if the interpretation by PR that both transitions originate from the segmental α-relaxation is correct. However, these techniques had found only a *single* glass transition at $T_g(h)$ approximately the same as $T_{gl}(h)$ of the lower transition of PR. To make this clear, we consider the dielectric [7,8] and PCS [6] data of freestanding films of high MW atactic polystyrene. The measured dielectric $\tau_\alpha(T)$ of 40 nm film over some temperature range are shown together with the Vogel-Fulcher fit in Fig.1. The dielectric glass transition temperature $T_{gd}$ is determined by extrapolating the fit of $\tau_\alpha(T)$ down to 1000 s, and by definition that $\tau_\alpha(T_{gd})=1000$ s. The values of $1000/T_{gu}(h)$ and $1000/T_{gl}(h)$ from ellipsometry by PR [1] for film with $h\approx40$ nm are represented by large close square and open circle respectively. The $T_{gl}(h)$ of the lower transition by ellipsometry from PR is in approximate agreement with the dielectric $T_{gd}$, verifying that the segmental α-relaxation is the mechanism giving rise to the lower transition of PR. Shown also in Fig.1 are data of $\tau_\alpha(T)$ of the freestanding 22 nm PS film, with $M_w$=767,000 g/mol and $M_w/M_n$=1.11, obtained by Forrest et al. [6] using photon correlation spectroscopy (PCS). This technique is also mainly sensitive to the segmental α-relaxation [10]. The $\tau_\alpha(T)$ data of the freestanding 22 nm PS film from PCS are sparse, and no VFT fit is made but the $\tau_\alpha(T)$ values are quite long already for making an estimate of the possible location of the PCS $T_{gPCS}$ in the figure, which is also in rough agreement with $T_{gl}(h)$ from PR by ellipsometry. This is another support of the segmental α-relaxation triggering the lower and genuine glass transition.

However, dielectric and PCS relaxation measurements found no trace of the upper transition at $T_{gu}$. We make this clear by the inset of Fig.1 where the isochronal dielectric loss spectra at $f = 4.5$ Hz for $h$= 40, 120, 225 nm and the bulk PS all exhibits only one peak corresponding to the lower transition at $T_{gl}$, but no intensity or peak at higher temperature that can be identified with the ellipsometry upper transition at $T_{gu}$, the position of which on the *x*-axis is indicated by the arrow. According to PR [1,9] the upper transition is the purported glass transition executed by the segmental α-relaxation, and this occurs in ~90% of the material. This hypothesis would lead anyone to expect the presence of another dielectric loss peak at higher temperature $T_{gu}$ with dielectric strength about 9 times higher than the lone loss peak observed [7,8] at $T_{gl}$, and identified with the lower transition. The dielectric spectra of Rotella et al. of the freestanding film with $h\approx33$ nm also show no evidence of the expected huge dielectric loss peak corresponding to the ellipsometry upper transition at $T_{gu}$, indicated by the inverted filled triangle in Fig.1.

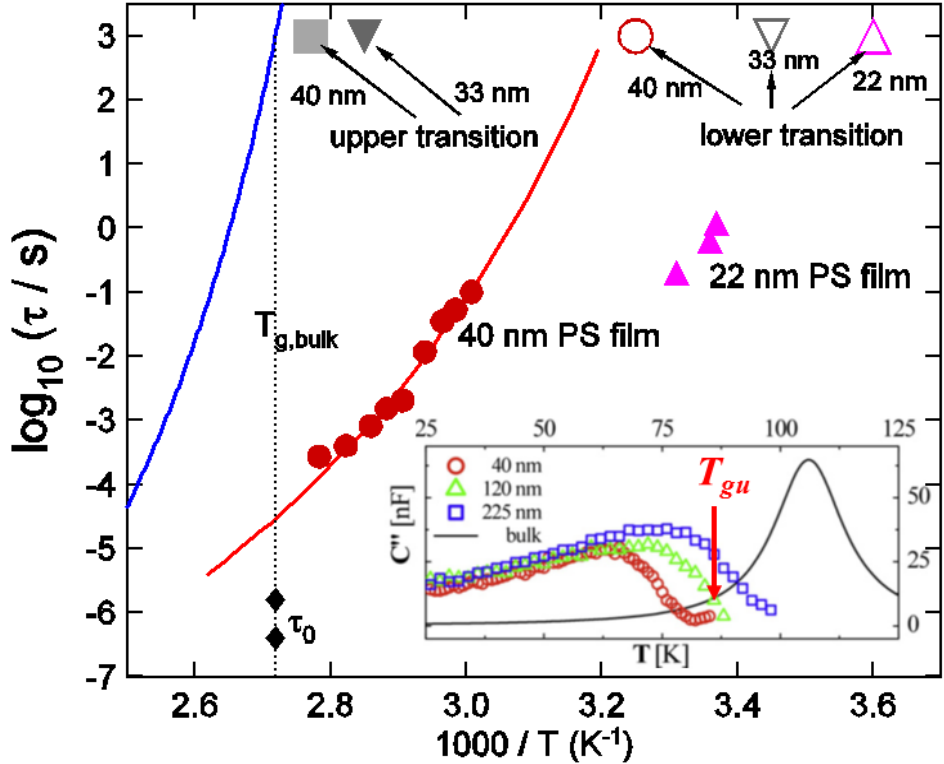

**Figure 1.** The segmental relaxation time, $\tau_\alpha(T)$, measured by dielectric relaxation on a 40 nm freestanding PS film (red filled circles, data from Rotella et al. [7]), and by PCS on a 22 nm freestanding PS film (magenta filled triangles, data from Forrest et al. [6]). The red line is the VFT fit to the dielectric $\tau_\alpha(T)$, The large open red circle and the two triangles (one is inverted) on the right have coordinates $(1000/T_{gl}, 3)$ estimated for the lower transition from ellipsometry measurements by Pye and Roth [1] for film thickness that is approximately the same as $h=$ 40, 33 and 22 nm respectively. The large closed square and inverted triangle on the left have coordinates $(1000/T_{gu}, 3)$ for the upper transition estimated from ellipsometry measurements by Pye and Roth for the same $h$ values. The blue line is the VFT fit of $\tau_\alpha(T)$ of bulk PS. The black filled diamonds are the primitive relaxation times of bulk PS calculated by the CM equation. The inset is the isochronal representation ($f = 4.5$ Hz) of the imaginary part of the complex capacitance for freely standing films of PS of different thickness. A peak of arbitrary intensity but the same peak maximum and full width at half height as for bulk samples has been added (black curve) for comparison. The arrow indicates $T_{gu}$. Inset reproduced from Ref.[8] by permission.

Thus, the dielectric relaxation data of freestanding PS thin films clearly contradict the interpretation of PR [1,9] that the upper transition originates from glass transition involving the segmental α-relaxation in ~90% of the PS film.

In contrast, the dielectric data are consistent with our identification of the upper transition with the sub-Rouse modes [11] because these modes in PS have negligible dielectric strength (molecular dipole is perpendicular to macromolecular chain) and cannot be resolved by dielectric spectroscopy [12]. The larger breadth of the upper transition observed by ellipsometry than the lower transition is consistent with the larger span of the sub-Rouse modes than the segmental α-relaxation in the compliance or the viscoelastic retardation spectrum of PS shown in Fig.1 of Ref.[13]. The fact that $T_{gu}$ does not depend on MW is also consistent with the sub-Rouse modes because they are independent of MW in high MW polystyrenes.

There is independent support of the enhancement of mobility of the sub-Rouse modes in PS thin films from biaxial creep compliance measurements by McKenna and co-workers [14] as explained by Ref.[13], and confirmed by more creep compliance measurements of thin films of a variety of polymers [15,16].

The experimental facts indicating that the sub-Rouse modes are coupled to density and pressure sensitive were presented in Ref.[17]. On lowering temperature past $T_{gu}$, the relaxation times of the sub-Rouse modes become too long to maintain their Vogel-Fulcher-Tammann dependence at equilibrium above $T_{gu}$. Hence the sub-Rouse modes, together with the density they coupled to, fall out of equilibrium and exhibit the upper transition observed by ellipsometry, in analogy to the segmental α-relaxation modes falling out of equilibrium at $T_{gl}$ of the lower transition. Again due to coupling of the sub-Rouse modes to density, physical aging performed at temperature below $T_{gu}$ in the 'glassy' state of the sub-Rouse modes will induce increase of their relaxation times and concomitantly densification of the upper 'glassy' state, as observed by PR in Ref.[9].

Besides relating their physical aging experiment to our paper [10], PR [9] also made a misguided remark on the mechanisms we considered in speeding up the segmental α-relaxation in ultrathin films. It is necessary to answer this remark made by PR [9] on our paper [10], although it is irrelevant to main point of whether the upper transition comes from the sub-Rouse modes or the segmental α-relaxation. Their remark is reproduced as follows: "*Furthermore, the main mechanism by which Prevosto et al. propose that the a-relaxation exhibits an increase in mobility, and concomitant decrease in the lower $T_g$, is due to chain orientation as the film thickness becomes smaller than the radius of gyration Rg of the polymer.[80,81,113] Formally the Coupling Model incorporates three factors by which the dynamics can be altered in confined polymer films: finite size effect, presence of free surfaces, and induced orientation of the chains. Refs. 80, 81, and 113 argue that chain orientation effects are the primary cause of the MW dependent lower $T_g(h)$ transition observed in high MW free-standing films.*" We are mystyfied by the last remark of "*Refs. 80, 81, and 113 argue that chain orientation effects are the primary cause of the MW dependent lower $T_g(h)$ transition observed in high MW free-standing films*."

No place in Ref. [80] by Prevosto et al. cited by PR [9] that we have said *chain orientation effects are the primary cause of the MW dependent lower $T_g(h)$ transition.* On the contrary, Prevosto et al. made two statments that show unequivocally the remark by PR is misguided. The belief by us that the free surfaces being a important if not primary cause was made in one statement "*For thin freestanding film of thickness h, the surfaces effectively mitigate the* *intermolecular coupling and cause reductions of $n_α$ and $n_{sR}$, which in turn shifts $τ_α(h)$ and $τ_{sR}(h)$ to shorter times in accordance with the CM.*" The point that the free surfaces and chain orientations are both considered together as causes was made in the other statement "*For freestanding polymer thin films of very high molecular weights, it was proposed that there are two factors acting together to reduce the coupling parameter $n_α$ and hence the glass transition temperature. One is the induced orientations of the polymer chains when their average end-to-end distance becomes comparable to the film thickness, and the other is the decrease of the cooperative length scale particularly for chain segments at and near the surfaces.[53,55–58]*".

In Ref.[81] cited by PR [9], we wrote "*When the thickness h of nanobubble-inflated film is reduced, several factors can alter the dynamics of the segmental αrelaxation than those found in the bulk polymer. One factor arises from h becoming comparable or less than the length scale of the cooperative segmental α-relaxation, together with the presence of the more mobile molecules at the free surfaces. The other factor is the induced polymer chain orientations in ultrathin polymer films.*"

In Ref.[113] cited by PR [9], we made several statements stressing the importance of the free surfaces in addition to chain orientations in thin films of high MW polystyrene. One example is the statement: "*The largest decrease of n from its bulk value occurs at the free surface and the change diminishes continuously when going towards the center of the film.*"

From these statements we made in Refs. 80, 81, and 113 of PR [9] as well as in Ref.18 in this Comment, we certainly have considered the important effect of the free surfaces in increasing the mobility of the segmental α-relaxation and lowering glass transition temperature in all our previous publications. In addition we consider the induced orientation effect when $h$ becomes comparable or less than the radius of gyration of the high MW polymer. This effect possibly can explain the large MW-dependence of $T_{gl}(h)$ of high MW PS and much less or lack of it in low MW PS [2,4]. Thus we are baffled by the misguided remark made by PR [9].

Moreover, it is interesting to note that in Ref.[1] PR wrote "*At present, we cannot provide information as to the location of this faster population within the film, but the cause of this very fast population is related to the chain connectivity of the polymer resulting in faster dynamics for higher MWs.*". Chain connectivity is the source of induced chain orientations in ultrathin films of high MW PS according to us. Thus in a way PR was considering implicitly the possible effect of induced chain orientation, that we suggested as a mechanism in high MW PS ultrathin films in addition to surface and finite size effects.